\documentstyle[eqsecnum,prd,aps,floats,graphicx]{revtex}
\draft

\newcommand{\square}{\kern1pt\vbox{\hrule height  1.2pt\hbox{\vrule
width 1.2pt\hskip 3pt
\vbox{\vskip 6pt}\hskip  3pt\vrule width 0.6pt}\hrule height
0.6pt}\kern1pt}
\begin{document}

\preprint{hep-th/0008188 v2}
\draft

% UNCOMMENT FOR TWO-COLUMN MODE
% \twocolumn[\hsize\textwidth\columnwidth\hsize\csname
% @twocolumnfalse\endcsname

\title{Dilaton-gravity on the brane}

\author{Kei-ichi Maeda$^{1,2}$ and David Wands$^3$}
\address{$^1$Department of Physics, Waseda University, Okubo 3-4-1,
Shinjuku, Tokyo 169-8555, Japan\\[-.7em]~}
\address{$^2$ Advanced Research Institute for Science and Engineering, 
Waseda University, Shinjuku, Tokyo 169-8555, Japan
\\[-.7em]~}
\address{$^3$Relativity and Cosmology Group,
School of Computer Science and Mathematics,\\ University of
Portsmouth, Portsmouth~PO1~2EG, United Kingdom\\[-.5em]~}

%\date{\today}

\maketitle

%======================================%
%<<<<<<<<<<<<< ABSTRACT >>>>>>>>>>>>>>>% 
%======================================%

\begin{abstract}
We derive the four-dimensional field equations for the induced metric
and scalar field on the world-volume of a 3-brane in a
five-dimensional bulk with Einstein gravity plus a self-interacting
scalar field. We calculate the effective four-dimensional
gravitational constant and cosmological constant for arbitrary forms
of the brane tension and self-interaction potential for the scalar
field in the bulk. In addition to the canonical energy-momentum tensor
for the scalar field and ordinary matter on the brane, the effective
four-dimensional Einstein equations include terms due to the scalar field 
and gravitational waves in the bulk.  We present solutions corresponding
to static Minkowski brane worlds and also dynamical
Friedmann-Robertson-Walker brane world cosmologies. We discuss the
induced coupling of the scalar field to ordinary matter on the brane.
\end{abstract}

\pacs{04.50.+h, 98.80.Cq~ \hfill hep-th/0008188 v2}

% UNCOMMENT FOR TWO-COLUMN MODE
% \vskip2pc]

\vskip1cm
\section{Introduction}

There has been tremendous interest over the last year in schemes for
dimensional reduction where ordinary matter fields are confined to
a lower-dimensional hypersurface, while only gravitational fields
propagate throughout all of spacetime. Such speculation has
been inspired by recent developments in string/M-theory where
gauge-fields can be associated with the degrees of freedom of open
strings whose end-points live on a D-brane hypersurface, while
gravitational fields associated with closed string excitations, can
propagate in the bulk~\cite{String,Polchinski,Witten}. In particular
Randall and Sundrum~\cite{RS2} proposed a static model where
four-dimensional Newtonian gravity is recovered at low energies when
considering linear perturbations about a Minkowski brane embedded in
five-dimensional anti-de Sitter spacetime (AdS$_5$).

Randall and Sundrum~\cite{RS2} considered pure 5-D Einstein gravity
in the bulk, sourced only by a cosmological constant, whereas in a
string theory or M-theory context one would also expect scalar fields,
associated with the many moduli fields, in the gravitational sector,
which should, in principle, also be allowed to propagate in the
bulk~\cite{Lukas,compact,duff}. For example, Lukas, Ovrut and
Waldram~\cite{Lukas} have derived an effective five-dimensional action
from a dimensional reduction from 11-dimensional M-theory which
contains scalar fields in the bulk which correspond to moduli
associated with the compactification of six-dimensions on a Calabi-Yau
space.

A particularly elegant way to study the effective four-dimensional
gravity induced in so-called `brane world' models was proposed by
Shiromizu, Maeda and Sasaki~\cite{ours}. They derived the effective
`Einstein equations' for the induced 4-D metric obtained by projecting
the five-dimensional metric onto the brane world-volume.  This
approach yields the most general form of the four-dimensional
gravitational field equations for a brane world observer whatever the
form of the bulk metric, in contrast to the usual Kaluza-Klein style
dimensional reduction which relies on taking a particular form for the
bulk metric in order to integrate over the fifth dimension.  The price
to be paid for such generality, is that the brane world observer may
be subject to influences from the bulk, in particular gravitational
waves, which are not constrained by local quantities, i.e., the set of
four-dimensional equations does not in general form a closed system.
Nonetheless, when the brane is located at an orbifold fixed point
under $Z_2$ symmetry (a ``brane at the end of the world'') the
energy-momentum tensor on the brane is sufficient to determine the
extrinsic curvature of the brane, and together with the local induced
metric, this strongly constrains the brane world gravity.  In the
absence of bulk matter, the effect of the bulk gravity can be
described by the projected five-dimensional Weyl tensor. In many
circumstances (e.g., AdS$_5$ bulk) this can be consistently set to
zero.

In this paper we will extend the work of Shiromizu, Maeda and
Sasaki~\cite{ours} to consider a simple dilaton-gravity theory with
scalar and tensor fields which propagate in the bulk, with
self-interaction potentials for the scalar field in the bulk and on
the brane. We will give the general form for the effective scalar
field and Einstein equations on the brane, in particular giving
expressions for the effective gravitational Newton's constant, and the
cosmological constant on the brane, both of which become functions of
the local scalar field. We also show how to recover previously found
solutions for static Minkowski brane worlds and dynamical
Friedmann-Robertson-Walker (FRW) cosmologies.

\section{The four-dimensional effective equations}

Suppose that the $i$-th brane world is located in the five-dimensional
spacetime as
\begin{eqnarray}
Y_{(i)}(X^A) = 0, 
\end{eqnarray}
where $X^A ~(A=0,1,2,3,5)$ are the 5-dimensional coordinates.
The effective action in 5-dimensional spacetime is then
% XXXX brane extrinsic curavture added in response to Harvey Reall's comment
\begin{eqnarray}
\label{action}
S = \int d^5X \sqrt{-{}^{(5)}g} \left[
{1 \over 2 \kappa_5^2} {}^{(5)}R -{1 \over 2} (\nabla \Phi)^2 -
{}^{(5)}\Lambda(\Phi) \right]
 +
\sum_i \int_{Y_{(i)}=0} d^4 x_{(i)} \sqrt{-g_{(i)}} 
\left[  {1\over\kappa_5^2} K_{(i)}^\pm
+ L_{(i)}^{\rm matter} -  \lambda_{(i)}(\Phi)
\right] ,
\end{eqnarray}
where $x_{(i)}^\mu ~(\mu=0,1,2,3) $ are the induced 4-dimensional 
brane world coordinates on the $i$-th brane
% XXXX
and ${}^{(5)}R$ is the 5-D intrinsic curvature in the bulk and
$K_{(i)}^\pm$ the extrinsic curvature on either side of the
brane~\cite{GH,ChaRea99}.
%
%==========================%

The five-dimensional Einstein equations in the bulk are
\begin{eqnarray}
{}^{(5)}G_{AB} = \kappa_5^2 {}^{(5)}T_{AB} \equiv
\kappa_5^2 \left[ \nabla_A\Phi \nabla_B \Phi - 
{}^{(5)}g_{AB} \left( {1 \over 2} (\nabla \Phi)^2 +
 {}^{(5)}\Lambda \right)
\right]
\label{bulkeom}
\end{eqnarray}
and the equation of motion for a dilaton $\Phi$ is 
\begin{eqnarray}
{}^{(5)}\square \Phi - {d {}^{(5)}\Lambda \over d \Phi}
 -\sum_i {d \lambda_{(i)} \over d\Phi} \delta (Y_i)
 =
\sum_i {1\over \Omega_{(i)}} {d \Omega_{(i)} \over d\Phi}\, \tau_{(i)}\, 
\delta (Y_i) \, ,
\end{eqnarray}
where $\tau_{(i)}$ is the trace of the energy-momentum tensor for
matter on the $i$-th brane.

This term on the right-hand-side of the equation arises because, in a
dilaton-gravity theory, one should allow for the possibility that
matter on the $i$-th brane is minimally coupled, not with respect to the
original Einstein metric, but with respect to a conformally related
metric
\begin{equation}
\label{CTmetric}
{}^{(5)}\tilde{g}_{AB} = \Omega_{(i)}^2(\Phi)\, {}^{(5)}g_{AB} \,.
\end{equation}
% XXXX comments added in response to Matt Visser question
Using the definition of the energy-momentum tensor
\begin{equation}
\tau_{(i)~\mu\nu} \equiv
-\frac{2}{\sqrt{-g_{(i)}}} \frac{\delta}{\delta
  g_{(i)}^{~\mu\nu}} \left( \sqrt{-g_{(i)}} L_{(i)}^{\rm matter} \right)
\end{equation}
we have
\begin{equation}
\frac{\delta}{\delta \Phi}
 \left( \sqrt{-g_{(i)}} L_{(i)}^{\rm matter} \right)
 = \frac{\partial
  \tilde{g}_{(i)}^{~\mu\nu}}{\partial \Phi} \frac{\delta}{\delta
  \tilde{g}_{(i)}^{~\mu\nu}} \left( \sqrt{-g_{(i)}}
L_{(i)}^{\rm matter}
\right)
 = \sqrt{-g_{(i)}} 
   \frac{1}{\Omega_{(i)}} \frac{d\Omega_{(i)}}{d\Phi}
   \, \tau_{(i)}
\end{equation}
which gives the right-hand-side in our Eq.(2.4).

To obtain the basic equations in one of the brane worlds, we project
the variables onto the four-dimensional brane world as in
Ref.\cite{ours}.  We shall assume that our four-dimensional world
($i=1$) is described by a domain wall (3-brane) $(M,g_{\mu\nu})$ at
$Y_1=0$, which as an orbifold fixed point under $Z_2$ reflection
symmetry, in a five-dimensional spacetime $({\cal M},{}^{(5)}g_{AB})$
and henceforth drop the index, $i$.
The induced 4D metric is obtained by projecting the five-dimensional
metric onto the brane world-volume:
\begin{equation}
g_{AB} = {}^{(5)}g_{AB} - n_{A}n_{B} \,,
\end{equation}
where $n_A$ is the spacelike unit-vector field normal to the brane. 
Using the Gauss and Codazzi equations, we obtain the induced
four-dimensional Einstein equations as~\cite{ours}
\begin{equation}
{}^{(4)}G_{\mu\nu}
 = {2 \kappa_5^2 \over 3}\left[{}^{(5)}T_{AB}
g^{~A}_{\mu}  g^{~B}_{\nu}  
+\left({}^{(5)}T_{AB}n^A n^B-{1 \over
4}{}^{(5)}T^C_{~C}\right)
 g_{\mu\nu} \right] 
+ KK_{\mu\nu} 
-K^{~\sigma}_{\mu}K_{\nu\sigma} -{1 \over 2}g_{\mu\nu}
  \left(K^2-K^{\alpha\beta}K_{\alpha\beta}\right) - E_{\mu\nu}, 
\label{4dEinstein}
\end{equation}
where the extrinsic curvature of $M$ is denoted by $K_{\mu\nu}=
g_\mu^{~C} g_\nu^{~D} \nabla_C n_D$, and $K=K^\mu_{~\mu}$ is its
trace.  
The effect of the bulk Weyl tensor is felt on the brane through the
symmetric, trace-free tensor~\cite{ours}
\begin{equation}
E_{\mu\nu} \equiv {}^{(5)}C^E_{~A F B}n_E n^F g_\mu^{~A} g_\nu^{~B} \,.
\label{Edef}
\end{equation}
The Codazzi equation yields an additional equation
\begin{equation}
D_\nu K^{~\nu}_\mu - D_\mu K = \kappa_5^2\, {}^{(5)}T_{B C}
n^C g_\mu^{~B} ,
\label{momentum}
\end{equation}
which turns out to be conservation of energy-momentum on the brane, as
we will see later, where $D_\mu$ is the covariant derivative with
respect to the induced metric $g_{\mu\nu}$.

The equation for $\phi$ can be reduced to
\begin{eqnarray}
\square \Phi -a_C \nabla^C \Phi + K {\rm \pounds}_n \Phi +
{\rm \pounds}_n^{~2}\Phi - {d {}^{(5)}\Lambda \over d \Phi} - {d \lambda
\over d\Phi}
\delta (Y_1) =
{1\over \Omega} {d \Omega \over d\Phi} \tau 
\delta (Y_1)  , 
\label{eq_Phi2}
\end{eqnarray}
where $\square \equiv D_\mu D^\mu , a^\mu \equiv n^\nu \nabla_\nu n^\mu$,
${\rm \pounds}_n$ is the Lie derivative in $n$ direction, and 
$\lambda \equiv \lambda_{(1)}$.

We can choose a coordinate $\chi$ such that the hypersurface $\chi=0$
coincides with our brane world ($Y_{(1)}=0$) and $n_A dX^A=d\chi$,
which is a condition on the coordinate in the direction of the extra
dimension. We assume this choice is possible at least in the
neighborhood of the brane.  This implies that
\begin{eqnarray}
a^A=0 , ~~{\rm \pounds}_n \Phi= \Phi', ~~ {\rm and } ~~{\rm \pounds}_n^2
\Phi=
\Phi'', 
\end{eqnarray}
where a prime denotes a differentiation with respect to $\chi$.
The assumption of $Z_2$ symmetry about the brane allows us to expand
the dilaton field $\Phi$ near the brane as
\begin{equation}
\Phi=\phi(x) + \Phi_1(x) |\chi| + {1 \over 2} \Phi_2(x) \chi^2
 + {\cal O}(\chi^3),
\end{equation}
and, inserting this into Eq. (\ref{eq_Phi2}), we find
the jump condition 
\begin{equation}
\Phi_1
 = {1 \over 2} \left( {d\lambda \over d \phi}
 + {1\over \Omega} {d \Omega \over d\phi} \tau \right),
\label{Phi_1}
\end{equation}
The equation for the dilaton, $\phi$, on the brane then becomes 
\begin{equation}
\label{dilatoneom}
\square \phi +
{K \over 2} \left( {d\lambda \over d \phi}
+ {1\over \Omega} {d \Omega \over d\phi} \tau \right)
- {d {}^{(5)}\Lambda \over d \phi} = - \Phi_2 \,.
\end{equation}

The junction condition for the extrinsic curvature due to the
energy-momentum tensor on the brane, $S_{\mu\nu}$, requires~\cite{israel}
\begin{equation}
\left[K_{\mu\nu}\right] = -\kappa_5^{~2} \left(
S_{\mu\nu}-{1\over 3} g_{\mu\nu} S\right) \,.
\end{equation}
With the ansatz of $Z_2$ symmetry, we have
$[K_{\mu\nu}]=K^+_{\mu\nu}-K^-_{\mu\nu}=2K_{\mu\nu}$ and hence we
obtain
\begin{equation}
K_{\mu\nu}
 = -{\kappa_5^{~2}\over 2} \left( S_{\mu\nu}-{1\over 3} g_{\mu\nu} S\right). 
\end{equation}
Since $\lambda$ is a four-dimensional vacuum energy, we can split
$S_{\mu\nu}$ as $S_{\mu\nu} = - \lambda g_{\mu\nu} + \tau_{\mu\nu}$,
where $\tau_{\mu\nu}$ is the energy-momentum tensor derived from the
matter Lagrangian on the brane,
and we then have 
\begin{eqnarray}
K_{\mu\nu} &=& -{1 \over 2} \kappa_5^{~2} \left(
{1\over 3} \lambda g_{\mu\nu} +\tau_{\mu\nu}-{1\over 3} g_{\mu\nu}
\tau \right), \nonumber \\
K &=& -{1 \over 6 }\kappa_5^{~2} \left(
4 \lambda - 
\tau \right) .
\label{extrinsic}
\end{eqnarray}

Finally then, the 4-dimensional effective Einstein
equations~(\ref{4dEinstein}) can be rewritten using the bulk
energy-momentum tensor, $^{(5)}T_{AB}(\Phi)$ given in
Eq.~(\ref{bulkeom}), the jump condition for the scalar field
derivative normal to the brane, Eq.~(\ref{Phi_1}), and the
expression for the brane extrinsic curvature in Eq.~(\ref{extrinsic}),
as
\begin{eqnarray}
{}^{(4)}G_{\mu\nu}=
{2\kappa_5^{~2}\over 3} \hat{T}_{\mu\nu}(\phi) 
+ \left[ -{}^{(4)}\Lambda +{\kappa_5^2\over16} \left( 2{d\lambda\over
d\phi} + {1\over\Omega}{d\Omega \over d\phi}\tau \right)
{1\over\Omega} {d\Omega \over d\phi} \tau \right] g_{\mu\nu} 
 + 8 \pi G_N(\phi) \tau_{\mu\nu}+\kappa_5^4\,\pi_{\mu\nu}
-E_{\mu\nu}\,, 
\label{eq:effective}
\end{eqnarray}
%
%==========================%
where
\begin{eqnarray}
\hat{T}_{\mu\nu}&=& D_\mu \phi D_\nu \phi -{ 5 \over 8} g_{\mu\nu}
(D\phi)^2 
\,,
\label{4d-dilaton}\\
{}^{(4)}\Lambda &=&\frac{1}{2}\kappa_5^2
\left[{}^{(5)}\Lambda +\frac{1}{6}\kappa_5^2\,\lambda^2
-{1 \over 8} \left({d\lambda \over d \phi}
\right)^2\right] \,,
\label{Lambda4}\\
8\pi G_N&=&{\kappa_5^4\over6}\lambda(\phi)\,,
\label{GNdef}\\
\pi_{\mu\nu}&=&
-\frac{1}{4} \tau_{\mu\alpha}\tau_\nu^{~\alpha}
+\frac{1}{12}\tau\tau_{\mu\nu}
+\frac{1}{8}g_{\mu\nu}\tau_{\alpha\beta}\tau^{\alpha\beta}-\frac{1}{24}
g_{\mu\nu}\tau^2\,.
\label{pidef}
\end{eqnarray}

This, of course, has many similarities with the effective Einstein
equations on the brane found previously for pure Einstein gravity in
the 5-D bulk~\cite{ours}.
The definition of $\pi_{\mu\nu}$, which represents the contribution of
terms quadratic in the 4-D energy-momentum tensor, has exactly the
same form and when $\lambda$ and $\Omega$ are independent of the
dilaton field, $\phi$, then the expressions for $^{(4)}\Lambda$ and
$G_N$ reduce to those found previously~\cite{ours}. 
More generally though, the effective Newton's constant on the brane,
$G_N$, becomes a function of the scalar field through the dependence
of the brane tension, $\lambda$, upon $\phi$.  Although $\lambda$
appears to act like an effective Brans-Dicke field, this is not easily
related to a four-dimensional Brans-Dicke type gravity theory (e.g.,
by a conformal transformation) due to scalar field kinetic terms and
other non-standard couplings to $\phi$ that appear in the effective
Einstein equations. For instance the trace of the energy-momentum tensor
 for
matter appears at linear order in the Einstein equations when
$d\Omega/d\phi\neq0$, which was recently noted for particular
cosmological solutions by Barger et al~\cite{BHLLM}.

Inserting the expression for the trace of the brane extrinsic
curvature given in Eq.~(\ref{extrinsic}) into the equation of motion
for the dilaton~(\ref{dilatoneom}) yields
\begin{eqnarray}
\label{effectivedil}
\square \phi
=
{d \over d \phi} \left( {}^{(5)}\Lambda + {\kappa_5^2\over6}\lambda^2
\right) 
- {\kappa_5^2 \over 12} \left[ {d\lambda \over d \phi}
  - (4\lambda-\tau) {1\over \Omega} {d \Omega \over d\phi} \right] \tau
- \Phi_2 \,.
\end{eqnarray}
The function $\Phi_2$ represents the effect of the evolution of the
scalar field in the bulk, upon the propagation of the field on the
brane. It cannot be determined locally by the field on the brane,
without solving the equations of motion in the bulk
% XXXX referee's comment
and there may be no consistent global solution for some choices of
$\Phi_2$. 
This represents a limitation in the uses of the induced
dilaton-gravity equations.  In the case of vacuum Einstein gravity in
the bulk the effect of bulk gravitational field is described solely by
the symmetric, trace-free tensor $E_{\mu\nu}$. Symmetry requirements,
together with the conservation of energy-momentum on the brane, may be
sufficient to completely determine the effect of $E_{\mu\nu}$ in some
situations (in particular, homogeneous
cosmologies~\cite{BDL,Flanagan,ours}) without knowing anything about
the solution for Einstein equations in the bulk. In the case of
scalar-tensor gravity in the bulk, $\Phi_2$ appears as a free-function
and will allow energy-momentum conservation for the scalar field on
the brane to be violated by transferring energy and/or momentum to the
bulk.

Using the bulk energy-momentum tensor, $^{(5)}T_{AB}(\Phi)$ given in
Eq.~(\ref{bulkeom}), and the expression for the brane extrinsic
curvature in Eq.~(\ref{extrinsic}) in the Codazzi equation
(\ref{momentum}), we find
\begin{equation}
{1 \over 2}\kappa_5^{~2} \left({d\lambda
\over d\phi} D_\mu \phi - D_\nu \tau_{\mu}^{~\nu}
 \right)
= \kappa_5^{~2} D_\mu \phi \cdot \Phi_1 \,.
\end{equation}
Using the jump condition for the scalar field, given in
Eq.~(\ref{Phi_1}), we obtain
\[
D_\nu \tau_{\mu}^{~\nu} = - {1\over\Omega} {d\Omega \over d\phi}
\tau D_\mu\phi , 
\]
which is consistent with conservation of energy-momentum for matter on
the brane, $\tilde{D}_\nu\tilde\tau_\mu^\nu=0$, with respect to the
conformally related metric $\tilde{g}_{\mu\nu}$ given by
Eq.~(\ref{CTmetric}), in which the conformally rescaled
energy-momentum tensor is
$\tilde\tau_{\mu\nu}=\tau_{\mu\nu}/\Omega^2$.

\section{The dilaton-vacuum brane world}
 
Henceforth we shall consider only dilaton-vacuum solutions on the brane
($\tau_{\mu\nu}=0$) in which case the effective Einstein equations
simplify considerably to give
\begin{eqnarray}
{}^{(4)}G_{\mu\nu}=
{2\kappa_5^{~2}\over 3} \hat{T}_{\mu\nu}(\phi) 
-{}^{(4)}\Lambda g_{\mu\nu} - E_{\mu\nu}\,.
\label{eq:effective2}
\end{eqnarray}
If we define a quantity $\Delta\Phi_2 $, such that
\begin{equation}
\label{eq:defQ}
\Delta\Phi_2 \equiv 
 \Phi_2
 - {1\over4} \, {d\lambda \over d\phi} \, {d^2\lambda \over d\phi^2} 
 \,,
\end{equation}
then we can rewrite the equation of motion of the dilaton on the
brane~(\ref{effectivedil}) as
\begin{equation}
\label{eq:dilaton2}
 \square \phi - {d U_{eff} \over d \phi} 
 =  -\Delta\Phi_2 \,,
\label{vacuum_dilaton2}
\end{equation}
where the effective potential for the scalar field on the brane is
\begin{equation}
\label{Ueff}
U_{eff} = {2\over\kappa_5^2} {}^{(4)}\Lambda_4
 = {}^{(5)}\Lambda +\frac{1}{6}\kappa_5^2\,\lambda^2
-{1 \over 8} \left( {d\lambda \over d \phi} \right)^2
\end{equation}
Introducing 
the canonical form of the energy-momentum tensor for the
scalar field on the brane as
\begin{eqnarray}
T_{\mu\nu}(\phi)  =D_\mu \phi D_\nu \phi -g_{\mu\nu}\left(
{ 1 \over 2} (D\phi)^2 +U_{eff} \right),
\end{eqnarray}
and a vector field $J_\mu = \Delta\Phi_2 \cdot D_\mu \phi$, we find that
Eq.(\ref{eq:dilaton2}) yields
\begin{eqnarray}
\label{defJ}
D^\nu T_{\mu\nu} = -J_\mu \,.
\end{eqnarray}
Thus we regard $J_\mu$ as the energy-momentum lost from the scalar
field on the brane to the bulk.

The four-dimensional Einstein equations 
Eq.(\ref{eq:effective2}) are now 
\begin{eqnarray}
{}^{(4)}G_{\mu\nu}= {2\kappa_5^{~2}\over 3} 
\left[T_{\mu\nu}(\phi) +\Delta T_{\mu\nu}(\phi)  \right] -E_{\mu\nu}\,, 
\label{eq:effective3}
\end{eqnarray}
where the contribution from the scalar field energy density in the
bulk is given by
\begin{equation}
\Delta T_{\mu\nu} = {1\over
4} g_{\mu\nu} \left( U_{eff} - {1\over 2}(D\phi)^2 \right) \,.
\end{equation}
This has the same equation of state as a cosmological constant, albeit
with a time-dependent value proportional to the Hamiltonian density
for the scalar field on the brane.

{}From the Einstein equations (\ref{eq:effective3}) and the Bianchi
identity, we have
\begin{eqnarray}
\label{DE}
D^\nu E_{\mu\nu} & = &{2 \kappa_5^{~2} \over 3} D^\nu
\left[ T_{\mu\nu}(\phi)+ \Delta T_{\mu\nu}(\phi)\right] 
\nonumber \\
&=& {2 \kappa_5^{~2} \over 3}\left[ D^\nu  \left(\Delta
T_{\mu\nu}\right)
 -  J_\mu \right].
\end{eqnarray}

\subsection{Minkowski brane worlds}

We shall now demonstrate that we can recover known static solutions
with four-dimensional Minkowski spacetime on the brane in the absence
of matter ($\tau_{\mu\nu}=0$) when there is no energy transfer between
bulk and brane ($J_\mu =0$) for particular forms of the bulk vacuum
energy, $^{(5)}\Lambda$, and brane tension, $\lambda$.

The Randall-Sundrum brane is a slice in five-dimensional anti-de
Sitter spacetime~\cite{RS1,RS2,MSM} with constant dilaton field (both in
the bulk, $\Phi_1=0$, $\Phi_2=0$, and on the brane, $D_\mu\phi=0$)
and vanishing Weyl curvature in the bulk, and hence $E_{\mu\nu}=0$ on
the brane.
We then recover four-dimensional Minkowski spacetime on the brane
[with $^{(4)}G_{\mu\nu}=0$ in Eq.~(\ref{eq:effective}) and
$\Box\phi=0$ in Eq.~(\ref{eq:dilaton2})] when the vacuum
energy in the bulk and brane tension are both independent of $\Phi$
and obey the Randall-Sundrum condition~\cite{RS1,RS2}
\begin{equation}
{}^{(4)}\Lambda = \frac{1}{2}\kappa_5^2
\left[{}^{(5)}\Lambda +\frac{1}{6}\kappa_5^2\,\lambda^2\right] = 0 \,.
\end{equation}
Indeed we see that any vacuum solution of four-dimensional general
relativity is a particular solution in this case.

In the case of a vanishing bulk cosmological constant,
$^{(5)}\Lambda=0$, one can also obtain Minkowski spacetime on the
brane with vanishing effective cosmological constant on the brane,
${}^{(4)}\Lambda=0$, even when the brane tension is non-zero. {}From
Eq.~(\ref{Lambda4}) we require
\begin{equation}
\label{self-tuning}
{\kappa_5^2\over6} \lambda^2
 = {1\over8} \left( {d\lambda \over d\phi} \right)^2
\end{equation}
and hence $\lambda\propto \exp(\pm2\kappa_5\phi/\sqrt{3})$, 
% XXXX comment of Harvey Reall
if this is to hold for all $\phi$.
For a coupling with this specific functional dependence upon the bulk
scalar, the vanishing of the four-dimensional cosmological constant is
independent of the actual amplitude of the brane tension. In
particular it may be independent of radiative corrections to the brane
tension due to matter fields on the brane offering a possible
resolution of the cosmological constant
problem~\cite{ADKS,KSS}. However the vanishing of ${}^{(4)}\Lambda$ is
sensitive to corrections to the vacuum energy in the bulk~\cite{FLLN}.

In the case of a non-vanishing bulk potential the existence of a Minkowski
brane requires
\begin{equation}
{}^{(5)}\Lambda + {\kappa_5^2\over6} \lambda^2
 = {1\over8} \left( {d\lambda \over d\phi} \right)^2
\label{zeroLambda4}
\end{equation}
An example is provided by the five-dimensional effective action
obtained by Lukas, Ovrut and Waldram~\cite{Lukas} from a dimensional
reduction of an eleven-dimensional Ho\v{r}ava-Witten model, in which
${}^{(5)}\Lambda =
(\alpha_0^2/6\kappa_5^{~2})e^{-2\sqrt{2}\kappa_5\Phi}$ and $\lambda =
(\sqrt{2}\alpha_0/\kappa_5^{~2})e^{-\sqrt{2}\kappa_5\Phi}$.  Note that
in this case the form of the action in Eq.~(\ref{action}) is invariant under
a global rescaling where $\Phi\to\Phi+$constant and
$\alpha_0\to\alpha_0\times$constant~\cite{ChaRea99}.  
We will return to this example when we discuss non-static solutions later.
For a constant scalar field to to be a solution to the equation of
motion for the scalar field on the brane, Eq.~(\ref{eq:dilaton2}),
we require in addition that $J_\mu=0$ and hence, from
Eq.~(\ref{eq:defQ}), 
\begin{equation}
\label{Phi_2}
\Phi_2 = {1\over4} \, {d\lambda \over d\phi} \, {d^2\lambda \over d\phi^2}
 \,.
\end{equation}
As $\Phi_2$ appears as a free function on the brane this is not a
restriction on the theory parameters, but one must solve the full
five-dimensional equations of motion to determine whether it is a
consistent solution in the bulk~\cite{Lukas}.

Quite generally, if one seeks a static solution in the bulk of the
form
\begin{equation}
ds_5^2 = e^{2A(\chi)} \eta_{\mu\nu} dx^\mu dx^\nu + d\chi^2 \,,
\end{equation}
where $e^{2A(y)}$ is commonly called the ``warp factor'', then the
five-dimensional Einstein equations~(\ref{bulkeom}) admit a
solution of the form~\cite{CLP,CEGH,BHLLM}
\begin{equation}
A' = -{1\over6} W(\Phi) \,, \qquad
\Phi' = {1\over 2\kappa_5^2} {dW\over d\Phi} \,.
\label{bulksolution}
\end{equation}
where the auxiliary field (or ``superpotential'') $W(\Phi)$ is
related to the bulk potential via the relation
\begin{equation}
\label{superpotential}
{1\over4\kappa_5^2} \left( {dW \over d\Phi} \right)^2 - {1\over 3} W^2 = 
 2\kappa_5^2 {}^{(5)} \Lambda
\end{equation}
The jump conditions at the brane, given in Eqs.~(\ref{extrinsic})
and~(\ref{Phi_1}) require in addition that
\begin{equation}
W(\phi) = \kappa_5^2 \lambda(\phi) \,.
\end{equation}
For a static brane solution with $\phi=\phi_c$ fixed then this
requires only that $W(\phi_c)=\kappa_5^2\lambda(\phi_c)$ at that
% XXXX reference suggested by Reall
particular value~\cite{Gubser}. 
But if this relation holds for all field values,
$\phi$, then inserting this relation into Eq.~(\ref{superpotential})
then shows that Eq.~(\ref{zeroLambda4}) for vanishing cosmological
constant on the brane is satisfied on any $\chi=$constant
hypersurface.  On the other hand substituting this relation into
Eq.~(\ref{bulksolution}) yields the required form for $\Phi_2$ given
in Eq.~(\ref{Phi_2}).

\subsection{FRW dilaton-vacuum brane cosmologies}

It is also possible to find Friedmann-Robertson-Walker (FRW)
cosmological solutions on the brane with time-dependent scale factor,
$a(t)$, and scalar field, $\phi(t)$, where $t$ is cosmic proper time,
if we know, or make some assumption about the energy transfer from
brane to bulk\footnote{
% XXXX in response referee's comment
  As remarked earlier, there is no guarantee that there exists a
  consistent global solution throughout the bulk for an arbitrary
  choice of $\Delta\Phi_2$ on the brane.},
$J_0= \dot{\phi}\,\Delta\Phi_2$.

The scalar field equation of motion (\ref{eq:dilaton2}) is
\begin{equation}
\ddot\phi + 3H \dot\phi = {dU_{eff}\over d\phi} + \Delta\Phi_2 \,,
\label{eq_scalar}
\end{equation}
where $H=\dot{a}/a$ is the Hubble expansion parameter,
while the Friedmann equation is 
\begin{equation}
H^2 +{k \over a^2}  = {2 \kappa_5^{~2} \over 9}\left(\rho_\phi + \Delta
\rho_\phi\right) + {1\over 3}E^0_{~0}
\,.
\label{eq_Friedmann}
\end{equation}
The energy density of the scalar field $\rho_\phi$ and the effect from
the scalar field in the bulk $\Delta\rho_\phi$ are given by
\begin{equation}
\rho_\phi=-T^0_{~0} = {1 \over 2} \dot{\phi}^2 + U_{eff} ~~~~{\rm and}
~~~~
\Delta\rho_\phi=- \Delta T^0_{~0}= -{1 \over 4} \rho_\phi
\label{eq_density}
\end{equation}
The equation for $E^0_{~0}$ is given from Eq.~(\ref{DE}) as
\begin{equation}
\dot{E^0_{~0}} + 4H E^0_{~0} = {2 \kappa_5^{~2} \over 3}
\left[ {1 \over 4} \dot{\rho}_\phi -J_0 \right],
\label{eq_E}
\end{equation}
where we have used $E_{~\mu}^\mu = 0$ and $\Delta
T_{\mu\nu}=(\rho_\phi/4)g_{\mu\nu}$.

\subsubsection{Solutions with energy conservation for dilaton field
on the brane}

When there is no energy transfer from brane to bulk, i.e. $J_0=0$, we
find a closed set of equations (\ref{eq_scalar}--\ref{eq_E}) for the
dilaton-vacuum universe, for a given $U_{eff}$.
Equations~(\ref{eq_scalar}) and (\ref{eq_E}) can then be integrated,
if the bulk and brane potentials obey the generalized Randall-Sundrum
condition given in Eq.~(\ref{zeroLambda4}), so that $U_{eff}=0$.
Equation~(\ref{eq_scalar}) can be simply integrated to give
\begin{equation}
\dot{\phi} = {C_\phi \over a^3}
\label{sol_phi}
\end{equation}
where $C_\phi $ is an integration constant, and Eq.~(\ref{eq_E}) becomes
\begin{equation}
\dot{E^0_{~0}} + 4H E^0_{~0} = {2 \kappa_5^{~2} \over
3} {d\over dt} \left({C_\phi^2\over 8 a^6}\right) \,.
\end{equation}
This in turn can be integrated to give
\begin{equation}
E^0_{~0} = {\kappa_5^{~2}
C_\phi^2 \over 4 a^6} + {{\cal E}_0 \over a^4},
\label{sol_E}
\end{equation}
where ${\cal E}_0$ is another integration constant.  Inserting
Eqs.~(\ref{sol_phi}) and (\ref{sol_E}) into Eq.~(\ref{eq_Friedmann}),
we find
\begin{equation}
H^2 +{k \over a^2}  = {\kappa_5^{~2}
C_\phi^2 \over 6 a^6} + {{\cal E}_0 \over 3 a^4}
\,.
\label{eq_Friedmann2}
\end{equation}
This is the same as the standard Friedmann equation with stiff matter,
with energy density proportional to $a^{-6}$, and radiation,
with energy density is proportional to $a^{-4}$.

The general solution with an initial singularity ($a=0$) can be
expressed in terms of the conformal time, $\eta=\int dt/a$,
as~\cite{MW95}
\begin{equation}
a^2
 = {\tau \left( \sqrt{6}\kappa_5 |C_\phi| + {\cal E}_0 \tau \right)\
    \over 3\left( 1+k\tau^2 \right)} \,,
\end{equation}
where 
\begin{equation}
\tau \equiv
\left\{
\begin{array}{cc}
|\eta| & {\rm for}\ k=0 \\
\tan|\eta| & {\rm for}\ k=+1 \\
\tanh|\eta| & {\rm for}\ k=-1 
\end{array}
\right.
\,,
\end{equation}
and
\begin{equation}
\sqrt{2\over 3}\kappa_5  (\phi-\phi_0)
 = \pm \ln \left| {{\cal E}_0 \tau \over \sqrt{6}\kappa_5 |C_\phi |
+ {\cal E}_0\tau} \right|
\,,
\end{equation}
for ${\cal E}_0\neq0$, or 
\begin{equation}
\sqrt{2\over 3}\kappa_5 (\phi-\phi_0)
 = \pm \ln \left| \tau \right|
\,,
\end{equation}
when ${\cal E}_0=0$, where $\pm $ corresponds to the sign of
$C_\phi$. The initial singularity appears at $\tau=0$. We also find a
big crunch at $ \tau=\tau_c\equiv
\sqrt{6}\kappa_5 |C_\phi| / |{\cal E}_0|
$ if ${\cal E}_0 <0$ for $k=0$ and 1 and if ${\cal E}_0
\leq -\sqrt{6}\kappa_5 |C_\phi|$ for $k=-1$.

Solutions with $C_\phi\neq0$ start expanding away from the initial
singularity with 
\begin{equation}
\label{stiffmatter}
a=\left({3 \over 2}\kappa_5^{~2}
C_\phi^{~2}\right)^{1/6} t^{1/3}
\end{equation}
This is just the standard cosmological solution for a spatially flat
universe with stiff matter. Since we have a massless dilaton field, we
naturally expect such a solution as a particular solution.  Spatially
flat models ($k=0$) with ${\cal E}_0>0$ then approach the standard 
radiation-dominated evolution with
\begin{equation}
a= \left({4{\cal E}_0 \over 3}\right)^{1/4} t^{1/2} \,,
\end{equation}
but if ${\cal E}_0<0$ then the universe will recollapse to a big
crunch when the conformal time reaches $\eta=\sqrt{6}\kappa_5
|C_\phi|/|{\cal E}_0|>0$. Although the similar big crunch appears for
$k=-1$ if ${\cal E}_0\leq -\sqrt{6}\kappa_5 |C_\phi|$, a non-singular
solution also exists in the same parameter range.  In particular, if
$C_\phi=0$, the open model with $k=-1$ and ${\cal E}_0<0$ is always
non-singular.

There is another exact solution for $k=-1$ when $ {\cal E}_0 \leq -
\sqrt{6}\kappa_5 |C_\phi|$, which is
\begin{equation}
a^2 = {a_0^{~2}\over
(1-\tau^2)}\left[ 1 -
\left({\tau \over \tau_*}\right)^2 \right]
\,,
\end{equation}
 and
\begin{equation}
 (\phi-\phi_0)
 = -3 C_\phi  \ln \left|{\tau_* -\tau \over \tau_* +\tau}\right|
\,,
\end{equation}
where
\begin{eqnarray}
\tau &=&\tanh |\eta| \nonumber\\
a_0^2 &=&{1\over 6}
\left[|{\cal
E}_0|+\sqrt{{\cal E}_0^2 -6\kappa_5^{~2}C_\phi^{~2}}\right]
\nonumber \\
\tau_*
&= &{1\over \sqrt{6}
\kappa_5 |C_\phi|}
\left[|{\cal
E}_0|+\sqrt{{\cal E}_0^2 -6\kappa_5^{~2}C_\phi^{~2}}\right] ~(\geq 1) .
\end{eqnarray}
This solution is non-singular: that is, the infinitely large universe
($a=\infty$) contracting from the past infinity ($t=-\infty$), bounces
at $t=0$ with a finite scale factor $a_0$, and then expands to
$a=\infty$ as $t\rightarrow
\infty$.

We summarize the asymptotic behaviors of the solutions in Tables 1 and
2 for the cases of $C_\phi \neq 0$ and $C_\phi =0$,
respectively.

\begin{table}[ht]
\caption[table1]{\footnotesize{The initial and final asymptotic
behaviors of the universe with
$C_\phi
\neq 0$ for each parameter range  of  $k$ and ${\cal E}_0$. 
There are two solutions for the case of $k=-1$ and ${\cal E}_0 \leq
-\sqrt{6}\kappa_5 |C_\phi|$.
[BC] means a big crunch, which occurs  at $t=t_c$. The universe
starts to expand from a big bang singularity and eventually collapses to
a big crunch singularity after reaching a maximum radius.  [SF] means a
singularity-free solution which has no initial or final singularity. Such
a universe may initially collapse from an infinite radius at $t=-\infty$,
and bounce at a finite radius
$a_0$, and then expand to an infinite radius as $t\rightarrow \infty$.
Note that 
$a=a_0$= constant if ${\cal E}_0 = -\sqrt{6}\kappa_5
|C_\phi|$.}}
\vskip .2cm
\begin{tabular}{|@{}c ||@{}c |@{}c |@{}r
@{}l|}
%\hline
&&&\multicolumn{2}{c|}{${\cal E}_0<0$}
\\
\cline{4-4}\cline{5-5}
 & \raisebox{1.5ex}[0pt]{${\cal E}_0 >0$} &
\raisebox{1.5ex}[0pt]{${\cal E}_0 =0$} &
\multicolumn{1}{c|}{~~~~~$-\sqrt{6}\kappa_5 |C_\phi|<{\cal E}_0 <0
$~~~~~}&\multicolumn{1}{c|}{~~~~~${\cal E}_0 \leq -\sqrt{6}\kappa_5
|C_\phi|$~~~~~}\\
\hline\hline
&&&\multicolumn{1}{c|}{~}
&\multicolumn{1}{c|}{ ~~~$t^{1/3} \Rightarrow
(t_c-t)^{1/3} {\rm ~[BC]}$~~~}
\\ \cline{5-5}
\raisebox{1.5ex}[0pt]{~~~~$ k=-1$~~~~}&\raisebox{1.5ex}[0pt]{$ 
t^{1/3} 
\Rightarrow   t$}&\raisebox{1.5ex}[0pt]{$  t^{1/3} 
\Rightarrow   t$} &\multicolumn{1}{c|}{\raisebox{1.5ex}[0pt]{$  t^{1/3} 
\Rightarrow   t$}} &\multicolumn{1}{c|}{$a_0 (
\neq 0$) $\Rightarrow t$ ~[SF]} \\
\hline
$k=0$ & $t^{1/3} \Rightarrow t^{1/2}$ & $t^{1/3}$ &
 \multicolumn{2}{c|}{$t^{1/3}
\Rightarrow (t_c-t)^{1/3} {\rm ~[BC]}$} \\
\hline
$k=1$ & $~~~~t^{1/3} \Rightarrow
(t_c-t)^{1/3} {\rm ~[BC]~~~}$ & ~~~$t^{1/3} \Rightarrow
(t_c-t)^{1/3} {\rm ~[BC]}~~~$ & \multicolumn{2}{c|}{$t^{1/3}
\Rightarrow (t_c-t)^{1/3} {\rm ~[BC]}$}
\\ 
%\hline
\end{tabular}
\vskip 1em
\end{table}

%\vskip .2cm
\begin{table}[ht]
\caption[table2]{\footnotesize{The initial and final asymptotic
behaviors of the universe with
$C_\phi = 0$ for each parameter range  of  $k$ and ${\cal E}_0$. [BC]
means a big crunch  at $t=t_c$.  [SF] means a singularity-free
solution.}}
\vskip .2cm
\begin{tabular}{ |   @{}c ||  @{}c |  @{}c |  @{}c| }
% \hline
 & ${\cal E}_0 >0$ &
${\cal E}_0 =0$ & ${\cal E}_0<0$ \\ \hline\hline
~~~~~~~$k=-1$~~~~~ & ~~~~~$t^{1/2} \Rightarrow t$~~~~~ &~~~~~ $t$ [Milne
universe] ~~~~~&
~~~~~~~~~~$a_0 (\neq 0)
\Rightarrow t $ ~[SF]~~~~~~~~~~
\\ \hline
$k=0$ & ~~~~~$t^{1/2}$~~~~~ & ~~~~~~~~~~$a_0$ [Minkowski space] ~~~~~~~~~~&
~~~~~no solution ~~~~~ \\
\hline
$k=1$ & ~~~~~~~~~~$t^{1/2} \Rightarrow (t_c -t)^{1/2}$ ~[BC]~~~~~~~~~~& no
solution  & no solution
\\ 
%\hline
\end{tabular}
\end{table}

\subsubsection{Solutions with time-dependent radion}

While the assumption of no energy transfer to or from the scalar field
on the brane gives simple cosmological solutions on the brane, these
differ from previously derived solutions for the full five-dimensional
dilaton-gravity 
% XXXX referee's comment
with $U_{eff}=0$ found by Lukas et al~\cite{Lukas}.
They considered a five-dimensional metric
\begin{equation}
ds_5^{~2} = - n^2dt^2 + a^2 d{\bf x}^2  + b^2 dy^2 
\end{equation}
where $n$, $a$ and $b$, and the scalar field $\Phi$ are all separable
functions of $t$ and $y$. In this case the five dimensional equations
can be related to a four-dimensional effective theory by a
Kaluza-Klein-type dimensional reduction where one integrates out the
$y$-dependence. Although the radion $b$ is non-minimally coupled to
the 4D metric and scalar, $\phi$, one can recover standard 4D Einstein
gravity with minimally coupled radion, by working in terms of the
conformally transformed time $\bar{t}$ and scale factor, $\bar{a}$,
where 
% XXXX response to referee's comment
(see, e.g., Ref.~\cite{LWC})
\begin{equation}
ds_5^{~2} = b^{-1} \left(-n^2 d\bar{t}^2 + \bar{a}^2 d{\bf x}^2\right) +
b^2 dy^2 \,.
\end{equation}
Conversely, working in the original frame, one finds an energy
transfer between the scalar field on the brane, $\phi(t)$, and the
radion field $b$, described by $J_0\propto (\dot{b}/b)\rho_\phi$. 
\footnote
{This ties in with our expectation that the energy density of the
scalar field in the bulk, (which contributes to the effective Einstein
equations on the brane through the non-local $\Delta T_{\mu\nu}$ and
$E_{\mu\nu}$ tensors) will change with the expansion/contraction of
the bulk metric.}
At the same time the expansion of the bulk metric is itself determined
by the local density.
In order to obtain separable solutions for the bulk metric, Lukas et
al~\cite{Lukas} require $\ln b\propto\Phi$ in the bulk, corresponding
to $J_0\propto \dot\phi^3$ on the brane.

%{\footnotesize{Table 2 :   The initial and final asymptotic
%behaviors of the universe with
%$C_\phi = 0$ for each parameter range  of  $k$ and ${\cal E}_0$. [BC]
%means a big crunch  at $t=t_c$.  [SF] means a singularity-free solution.}}
%
%\vskip 0.5cm

\begin{figure}[t]
\centering 
\includegraphics[height=6cm]{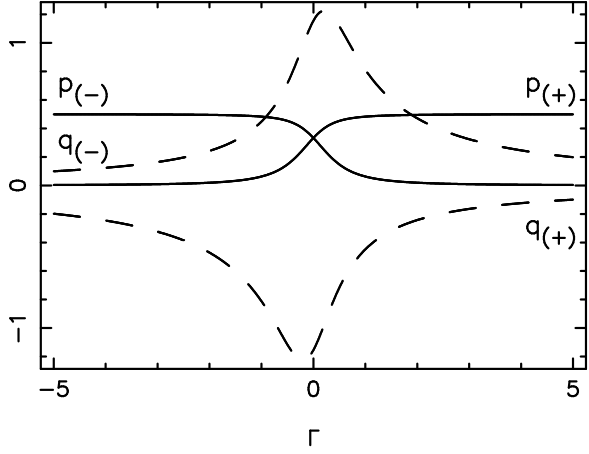}
\caption[pq]{\label{pq} Plot of the values of the exponents
$p_{(\pm)}$ (solid lines) and $q_{(\pm)}$ (dashed lines) given in
Eqs.~(\ref{sol:p}) and~(\ref{sol:q}).}
\end{figure}

If we describe the brane to bulk energy transfer by
\begin{equation}
J_0 = -\sqrt{2}\Gamma \kappa_5 \dot\phi^3 \,,
\end{equation}
where $\Gamma$ is a constant,
then Eq.~(\ref{eq_scalar}), with $dU_{eff}/d\phi=0$, 
can be integrated to give
\begin{equation}
\dot\phi = {C_\phi \over a^3 e^{\sqrt{2}\Gamma \kappa_5 \phi}} \,,
\end{equation}
which reduces to our previous solution given in Eq.~(\ref{sol_phi})
when $\Gamma=0$. The remaining equations (\ref{eq_E}) and
(\ref{eq_Friedmann}) can be integrated if we make the power-law
ansatz
\begin{equation}
a \propto |t|^p ~~~{\rm and }~~~~ 
e^{\sqrt{2}\kappa_5\phi} \propto |t|^q \,,
\label{sol_power-law}
\end{equation}
where the dilaton evolution Eq.~(\ref{eq_scalar}) requires $3p+\Gamma q=1$.
Equation (\ref{eq_E}) yields
\begin{equation}
{E}_0^0 =
- {(4p-1) q^2\over 8(2p-1)} |t|^{-2} +
{\cal E}_0  |t|^{-4p}.
\label{sol_E2}
\end{equation}
The second term in the right hand side is the ``dark radiation'' term,
which is proportional to $a^{-4}$.  For consistency with our power-law
ansatz, we require that ${\cal E}_0$ must be zero except for $p=1/2$.
Inserting the solution (\ref{sol_E2}) into 
% XXXX referee's comment
the modified Friedmann equation~(\ref{eq_Friedmann}) with $U_{eff}=0$ 
for $p\neq1/2$, we find $12p(2p-1) =
-q^2$. Combined with the requirement that $3p+\Gamma q=1$ this yields
the quadratic
\begin{equation}
3(3+8\Gamma^2) p^2 - 6(1+2\Gamma^2) p + 1 = 0 \,.
\end{equation}
We then find a one-parameter family of power-law solutions
(\ref{sol_power-law}) with exponents (see Figure~\ref{pq})
\begin{eqnarray}
\label{sol:p}
p &=& p_{(\pm)} ={3(2\Gamma^2+1)\pm2\sqrt{3}\Gamma\sqrt{3\Gamma^2+1}
\over 3(8\Gamma^2+3)}\\ 
\label{sol:q}
q &=& q_{(\pm)}
={2[\Gamma\mp\sqrt{3(3\Gamma^2+1)}]
\over 8\Gamma^2+3}.
\end{eqnarray}
For $\Gamma=1$ we recover the solutions of Lukas et
al~\cite{Lukas} with
\begin{equation}
p_{(\pm)}={3\over 11} \left(1\pm {4\sqrt{3} \over 9}\right)
~~~{\rm and }~~~~ q_{(\pm)}={2\over 11} \left(1\mp 2\sqrt{3}\right) \,.
\end{equation}
These solutions were generalised to spatially curved FRW models
including an additional bulk scalar field by Reall~\cite{Harvey}, and
for different potential exponents by Lidsey~\cite{Jim}. In the
notation of Ref.~\cite{ChaRea99} they are type II solutions.

The ``dynamical instability'' of the self-tuning potentials [obeying
the condition given in Eq.~(\ref{self-tuning})] reported in
Ref.~\cite{BCG} is recovered for $\Gamma=\pm\sqrt{1/24}$ where we
obtain $p=1/4$ and $q=\pm\sqrt{3/2}$. The 4D observer sees this as the
runaway behaviour of a massless scalar field in a contracting
cosmology for $t<0$ which is always losing energy to the bulk
($J_0<0$).

We recover the stiff-matter dominated solution given in
Eq.~(\ref{stiffmatter}) with $p_{(\pm)}=1/3, q_{(\pm)}=\mp 2/\sqrt{3}$
for $\Gamma=0$. Over the entire range $-\infty < \Gamma <\infty$,
$p_{(+)}$ changes monotonically from 0 to 1/2, while $q_{(+)}$ always
takes negative values from zero to zero, having its minimum value
$-\sqrt{6}/2$ at $\Gamma = -\sqrt{6}/12$.  Note that the solution with
$(p_{(-)}(\Gamma ), q_{(-)}(\Gamma ))$ is the same as that obtained by
changing the sign of $\Gamma$, i.e. the solution with
$(p_{(+)}(-\Gamma ), - q_{(+)}(-\Gamma ))$.

\section{Conclusion and Remarks}

In this paper we have derived the equations of motion for the induced
metric and the scalar field on a four-dimensional brane world embedded
in five-dimensional Einstein gravity plus a scalar field, $\Phi$, with
self-interaction potential energy ${}^{(5)}\Lambda(\Phi)$ and
brane tension
$\lambda(\Phi)$. In the case where $\Phi=$~constant we recover the
modified Einstein equations first derived in Ref.~\cite{ours}. 

More generally, we find that the induced cosmological constant on the
brane is given by the sum of three parts (i) the bulk cosmological
constant, ${}^{(5)}\Lambda$, (ii) the brane tension, which yields
extrinsic curvature of the brane, $\lambda^2$, and, (iii) the first
derivative of the brane tension, which leads to a discontinuity in the
scalar field gradient normal to the brane. The resulting effective
cosmological constant is given in Eq.~(\ref{Ueff}) as
\begin{equation}
{}^{(4)}\Lambda
 = {\kappa_5^2\over2} \left[
{}^{(5)}\Lambda +\frac{1}{6}\kappa_5^2\,\lambda^2
-{1 \over 8} \left( {d\lambda \over d \phi} \right)^2 \right] \,.
\end{equation}
Setting this to zero generalises the Randall-Sundrum
condition~\cite{RS1,RS2} (recovered for $d\lambda/d\phi=0$), needed to
obtain 4D Minkowski space-time solutions on the brane. This is
automatically satisfied for all values of the scalar-field in
supergravity theories~\cite{Lukas,CLP} where the bulk and brane
potentials are derived from the same superpotential. A special case is
provided by the self-tuning potentials~\cite{ADKS,KSS} for which
${}^{(5)}\Lambda=0$ and $\lambda\propto e^{\pm2\phi/\sqrt{3}}$.
It is interesting to note that a static solution does not require a
minimum for either ${}^{(5)}\Lambda$ or $\lambda$, but only that the
induced ${}^{(4)}\Lambda$ has a stationary value.

In addition to the canonically defined energy-momentum tensor for the
scalar field on the brane, the energy density of the scalar and
gravitational waves in the bulk also contribute to the 4D spacetime
curvature. The energy transfer between brane and bulk as seen by the
4D observer is parameterised by the 4-vector
$J_\mu=(D_\mu\phi)\Delta\Phi_2$, which cannot be determined by the 4D
equations but must be determined from the full 5D
solution. Nonetheless we are able to recover known FRW cosmological
solutions with time-dependent $\phi(t)$ by parameterising
$J_0\propto\dot\phi^3$.

In this paper we have focussed upon dilaton-vacuum solutions on the
brane, but in the presence of ordinary matter, the brane tension,
$\lambda(\phi)$, determines the strength of the gravitational
(Newton's) constant on the brane, $G_N\propto \lambda$.  The evolution
equation for the scalar field then couples the field on the brane to
the trace of the energy-momentum tensor, $\tau$, for ordinary matter on the
brane. {}From Eqs.~(\ref{effectivedil}) and~(\ref{defJ}) we obtain
\begin{equation}
J_\mu = (D_\mu\phi) \left\{ 
\Delta \Phi_2 
+ {\kappa_5^2 \over 12} \left[ {d\lambda \over d \phi}
  - (4\lambda-\tau) {1\over \Omega} {d \Omega \over d\phi} \right] \tau
\right\} \,,
\end{equation}
where we allow for the possibility that matter may be minimally
coupled in a conformally related metric
$\tilde{g}_{\mu\nu}=\Omega^2(\phi) g_{\mu\nu}$.  Such a picture has
important consequences for any attempt to understand the present value
of the 4D cosmological constant, and its relation to the 4D Planck
scale, in a brane world
context~\cite{ADKS,KSS,FLLN,CLP,CEGH,BHLLM,BCG,Verlinde}.

\bigskip

While this work was being completed we became aware of related work by
Barcel\'o and Visser~\cite{BV} and Mennim and Battye~\cite{MB}.

\acknowledgments

The authors are grateful to the organisers of the Structure Formation
in the Universe workshop at the Isaac Newton Institute for
Mathematical Sciences, Cambridge, where this work was begun.
KM is supported partially
by the Grant-in-Aid for Scientific Research Fund of the Ministry of
Education, Science and Culture (Specially Promoted Research No.\
08102010)  and by the Yamada foundation.
DW is supported by the Royal Society. 
% XXXX expanded list of acknowledgements
The authors thank Richard Battye, Roy Maartens, Andrew Mennim, Harvey
Reall and Matt Visser for useful comments and discussions.


\begin{thebibliography}{99}
\bibitem{String}
J. Polchinski, {\it String Theory I {\rm \&} II}
(Cambridge Univ. Press, Cambridge, 1998).
\bibitem{Polchinski}
J. Polchinski, Phys. Rev. Lett. {\bf 75}, 4724 (1995).
\bibitem{Witten}
P. Horava and E. Witten, Nucl. Phys. {\bf B460}, 506 (1996); 
ibid {\bf B475}, 94 (1996) 
\bibitem{RS2}
L.~Randall and R.~Sundrum,
%``An alternative to compactification,''
Phys.\ Rev.\ Lett.\  {\bf 83}, 4690 (1999)
[hep-th/9906064].
\bibitem{Lukas}
A. Lukas, B. A. Ovrut, K.S. Stelle and D. Waldram, Phys. Rev. {\bf
D59}, 086001 (1999) hep-th/9803235; A. Lukas, B. A. Ovrut and
D. Waldram, Phys. Rev. {\bf D60}, 086001 (1999) hep-th/9806002;
Phys. Rev. {\bf D61}, 023506 (2000) hep-th/9902071.
\bibitem{compact}
C.~S.~Chan, P.~L.~Paul and H.~Verlinde,
%``A note on warped string compactification,''
Nucl.\ Phys.\  {\bf B581}, 156 (2000)
[hep-th/0003236].
\bibitem{duff}
M. J. Duff, J. T. Liu and K. S. Stelle, 
%``A supersymmetric type IIB Randall-Sundrum scenario''
hep-th/0007120.
\bibitem{ours}
T. Shiromizu, K. Maeda and M. Sasaki, Phys. Rev. {\bf D62}, 024012 (2000)
 [gr-qc/9910076].
\bibitem{GH}
G.~W.~Gibbons and S.~W.~Hawking, Phys. Rev. {\bf D15}, 2752 (1977).
\bibitem{ChaRea99}
H.~A.~Chamblin and H.~S.~Reall,
%``Dynamic dilatonic domain walls,''
Nucl.\ Phys.\  {\bf B562}, 133 (1999)
[hep-th/9903225].
\bibitem{israel}
W. Israel, Nuovo Cim. {\bf 44B}, 1 (1966).
\bibitem{BHLLM}
V.~Barger, T.~Han, T.~Li, J.~D.~Lykken and D.~Marfatia,
%``Cosmology and hierarchy in stabilized warped brane models,''
hep-ph/0006275.
\bibitem{BDL}
P.~Bin\' etruy, C.~Deffayet and D.~Langlois, hep-th/9905012;
P.~Bin\' etruy, C.~Deffayet, U.~Ellwanger and D.~Langlois, hep-th/9910219.
\bibitem{Flanagan}
E.E. Flanagan, S.-H.H. Tye, and I. Wasserman, hep-ph/9910498.
\bibitem{RS1}
L.~Randall and R.~Sundrum,
%``A large mass hierarchy from a small extra dimension,''
Phys.\ Rev.\ Lett.\  {\bf 83}, 3370 (1999)
[hep-ph/9905221].
\bibitem{MSM}
S. Mukohyama, T. Shiromizu and K. Maeda, Phys. Rev. D {\bf 62}, 024028
(2000) [hep-th/9912287].
\bibitem{ADKS}
N.~Arkani-Hamed, S.~Dimopoulos, N.~Kaloper and R.~Sundrum,
%``A small cosmological constant from a large extra dimension,''
Phys.\ Lett.\  {\bf B480}, 193 (2000)
[hep-th/0001197].
\bibitem{KSS}
S.~Kachru, M.~Schulz and E.~Silverstein,
%``Self-tuning flat domain walls in 5d gravity and string theory,''
hep-th/0001206.
\bibitem{FLLN}
S.~Forste, Z.~Lalak, S.~Lavignac and H.~P.~Nilles,
%``The cosmological constant problem from a brane world perspective,''
hep-th/0006139.
\bibitem{CLP}
M.~Cvetic, H.~Lu and C.~N.~Pope,
%``Domain walls and massive gauged supergravity potentials,''
hep-th/0001002.
\bibitem{CEGH}
C.~Csaki, J.~Erlich, C.~Grojean and T.~Hollowood,
%``General properties of the self-tuning domain wall approach to the
%  cosmological constant problem,'' 
hep-th/0004133.
\bibitem{Gubser}
S.~S.~Gubser, hep-th/0002160.
\bibitem{MW95}
J.~P.~Mimoso and D.~Wands,
%``Massless fields in scalar - tensor cosmologies,''
Phys.\ Rev.\  {\bf D51}, 477 (1995)
[gr-qc/9405025].
\bibitem{LWC}
J.~E.~Lidsey, D.~Wands and E.~J.~Copeland,
%``Superstring cosmology,''
Phys.\ Rep.\ {\bf 337}, 343 (2000)
[hep-th/9909061].
\bibitem{Harvey}
H.~S.~Reall,
%``Open and closed cosmological solutions of Horava-Witten theory,''
Phys.\ Rev.\  {\bf D59}, 103506 (1999)
[hep-th/9809195].
\bibitem{Jim}
J.~E.~Lidsey,
%``Separable brane cosmologies in heterotic M-theory,''
Class.\ Quant.\ Grav.\  {\bf 17}, L39 (2000)
[gr-qc/9911066].
\bibitem{BCG}
P. Bin\' etruy, J. M. Cline and C. Grojean, hep-th/0007029.
%\bibitem{wald}
%R. M. Wald, {\it General Relativity},
%(Univ. Chicago Press, Chicago, 1984).
\bibitem{Verlinde}
E.~Verlinde and H.~Verlinde,
%``RG-flow, gravity and the cosmological constant,''
JHEP {\bf 0005}, 034 (2000)
[hep-th/9912018].
\bibitem{BV}
C. Barcel\' o and M. Visser, gr-qc/0008008.
\bibitem{MB}
A. Mennim and R. A. Battye, hep-th/0008192.
\end{thebibliography}
\end{document}